\documentclass[3p,10pt,fleqn]{elsarticle}

\usepackage{graphicx}
\usepackage[figuresright]{rotating}
\usepackage{bm,amsmath,amssymb}
\usepackage[mathscr]{eucal}
\usepackage{color}
\usepackage{caption}
\usepackage[normalem]{ulem}
\usepackage{setspace}
\usepackage{multirow}

\biboptions{square,comma,numbers,sort&compress}

\long\def\comment#1{ }
\newcommand{\eqn}[1]{Eq.~\eqref{#1}}
\newcommand{\beq}{\begin{eqnarray}}
  \newcommand{\eeq}{\end{eqnarray}}

\newcommand{\dif}{{\rm d}}
\newcommand{\rmd}{{\rm d}}
\newcommand{\rme}{{\rm e}}
\newcommand{\rmi}{{\rm i}}

\newcommand{\rmK}{{\rm K}}

\newcommand{\rmL}{{\rm L}}
\newcommand{\rmT}{{\rm T}}
\newcommand{\del}{\partial}

\newcommand{\order}[1]{\mcal{O}{(#1)}}
\newcommand{\mcal}{\mathcal}

\newcommand{\bk}{\bm{k}}

\newcommand{\bq}{\bm{q}}

\newcommand{\bx}{\bm{x}}
\newcommand{\by}{\bm{y}}

\newcommand{\bz}{\bm{z}}

\newcommand{\br}{\bm{r}}

\newcommand{\abar}{\bar{\alpha}_s}

\newcommand{\Nc}{N_{\rm c}}
\newcommand{\CF}{C_{\rm F}}
\newcommand{\Nf}{N_{\rm f}}

\newcommand{\minus}{\!-\!}

\newcommand{\kt}{k_\perp} 

\begin{document}

\begin{frontmatter}

\title{{\bf HERA data and collinearly-improved BK dynamics}}

\author[sac,edin]{B.~Duclou\'e}
\ead{bertrand.ducloue@ed.ac.uk}

\author[sac]{E.~Iancu\corref{cor1}}
\ead{edmond.iancu@ipht.fr}

\author[sac]{G.~Soyez}
\ead{gregory.soyez@ipht.fr}

\author[ect]{D.N.~Triantafyllopoulos}
\ead{trianta@ectstar.eu}

\address[sac]{Institut de physique th\'{e}orique, Universit\'{e} Paris Saclay, CNRS, CEA, F-91191 Gif-sur-Yvette, France}
\address[edin]{Higgs Centre for Theoretical Physics, University of Edinburgh, Peter
Guthrie Tait Road, Edinburgh EH9 3FD, UK}
\address[ect]{European Centre for Theoretical Studies in Nuclear Physics and Related Areas (ECT*)\\and Fondazione Bruno Kessler, Strada delle Tabarelle 286, I-38123 Villazzano (TN), Italy}

\cortext[cor1]{Corresponding author}

\begin{abstract}
  Within the framework of the dipole factorisation, we use a recent
  collinearly-improved version of the Balitsky-Kovchegov equation to
  fit the HERA data for inclusive deep inelastic scattering at small
  Bjorken $x$. The equation includes an all-order resummation of
  double and single transverse logarithms and running coupling
  corrections.
  Compared to similar equations previously proposed in the literature,
  this work makes a direct use of Bjorken $x$ as the rapidity scale
  for the evolution variable.
  We obtain excellent fits for reasonable values for the four fit
  parameters.
  We find that the fit quality improves when including resummation
  effects and a physically-motivated initial condition. In particular,
  the resummation of the DGLAP-like single transverse logarithms has a
  sizeable impact and allows one to extend the fit up to relatively
  large photon virtuality $Q^2$.
\end{abstract}


\end{frontmatter}

\section{Introduction}
\label{sect:intro}

Derived via systematic approximations within perturbative QCD, the Colour Glass Condensate 
(CGC) effective theory \cite{Iancu:2002tr,Iancu:2003xm,Gelis:2010nm,Lappi:2010ek,Kovchegov:2012mbw} is a powerful framework for computing high-energy processes in
the presence of non-linear effects associated with high parton densities. There are intense
ongoing efforts towards extending this effective theory to next-to-leading order (NLO) accuracy,
as required by realistic applications to phenomenology \cite{Balitsky:2006wa,Kovchegov:2006vj,Balitsky:2008zza,Balitsky:2013fea,Kovner:2013ona,Kovner:2014lca,Lublinsky:2016meo,Chirilli:2011km,Chirilli:2012jd,Iancu:2016vyg,Ducloue:2017dit,Beuf:2016wdz,Beuf:2017bpd,Roy:2019hwr}.
 These efforts refer both to the
Balitsky-JIMWLK equations  \cite{Balitsky:1995ub,JalilianMarian:1997jx,JalilianMarian:1997gr,Kovner:2000pt,Iancu:2000hn,Iancu:2001ad,Ferreiro:2001qy}, which govern the high-energy evolution of the scattering amplitudes, and to the impact factors, which represent cross-sections at relatively low energy. 
The first NLO results, obtained
more than a decade ago \cite{Balitsky:2006wa,Kovchegov:2006vj,Balitsky:2008zza},
refer to the Balitsky-Kovchegov (BK) equation~\cite{Balitsky:1995ub,Kovchegov:1999yj}. The latter is 
a non-linear equation emerging from the B-JIMWLK hierarchy in the
limit of a large number of colours ($N_c\to\infty$). It 
describes the evolution of the elastic scattering amplitude  
between a colour dipole and a dense hadronic target.
Via appropriate  factorisation schemes, like the ``dipole factorisation'' or the  ``hybrid factorisation''~\cite{Albacete:2013tpa},
the BK equation also governs the high-energy evolution
of the cross-sections for processes of phenomenological interest, like deep inelastic scattering (DIS) at small Bjorken $x$,
or forward particle production in proton-nucleus collisions.

A few years after the full NLO BK equation was first presented
\cite{Balitsky:2008zza}, its numerical study in~\cite{Lappi:2015fma}
showed that it is {\it unstable}, as  anticipated
in~\cite{Avsar:2011ds,Beuf:2014uia}. Similar problems had been
identified, and
cured~\cite{Kwiecinski:1997ee,Salam:1998tj,Ciafaloni:1998iv,Ciafaloni:1999yw,Ciafaloni:2003rd,Vera:2005jt},
for the NLO version of the BFKL
equation~\cite{Lipatov:1976zz,Kuraev:1977fs,Balitsky:1978ic} --- the
linearised version of the BK equation, valid for weak scattering.
The origin of this difficulty has been clearly identified, both
numerically~\cite{Lappi:2015fma} and
conceptually~\cite{Beuf:2014uia,Iancu:2015vea}: it is associated with
large and negative NLO corrections enhanced by a double
``anti-collinear'' logarithm, generated by integrating out gluon
emissions with small transverse momenta (see
Sect.~\ref{sect:evol}). Such double-logarithmic corrections spoil the
convergence of the fixed-order perturbative expansion of the
high-energy evolution, unless they are properly resummed to all
orders. Refs.~\cite{Beuf:2014uia,Iancu:2015vea} proposed two different
strategies for resumming this series of double-logarithmic corrections
to all orders.

At a first sight, these strategies seemed to be
successful, leading to stable evolution equations
\cite{Iancu:2015vea,Lappi:2016fmu} and allowing for good fits to the
small-$x$ HERA data~\cite{Iancu:2015joa,Albacete:2015xza}.
However, a recent study \cite{Ducloue:2019ezk} revealed some inconsistencies in
the original analyses in
\cite{Beuf:2014uia,Iancu:2015vea,Iancu:2015joa,Albacete:2015xza}.
In particular, 
there was a confusion concerning the meaning of the rapidity variable which plays the role
of the evolution time. The variable which {\em a priori} enters the perturbative calculations
at NLO~\cite{Balitsky:2008zza}
and in the resummed equations proposed in~\cite{Beuf:2014uia,Iancu:2015vea}
is the rapidity $Y\equiv\ln(s/Q_0^2)$ of the dipole projectile, with $s$
the centre-of-mass  energy squared and $Q_0$ a typical transverse momentum scale for the
target.
It is different from the rapidity $\eta\equiv\ln(1/x)$ of the hadronic
target, which is the variable used in DIS
(see Sect.~\ref{sect:evol} for details). When translated to this physical variable,
the results of the original resummations in~\cite{Beuf:2014uia,Iancu:2015vea} show a
strong scheme dependence, preventing any meaningful phenomenological applications~\cite{Ducloue:2019ezk}.
The success of the corresponding fits to the HERA
data~\cite{Iancu:2015joa,Albacete:2015xza} was rather fortuitous and can be attributed
to several factors:  \texttt{(i)} these fits blindly assumed the physical rapidity variable $\eta=\ln(1/x)$ (although this 
was inconsistent with the resummation scheme),  \texttt{(ii)} the
rapidity interval over which one can probe high-energy evolution
is relatively limited making it delicate to critically probe
the effects of resummation. In other words, even though the fits can
hint at an evolution being better than another (e.g.\ via a better
$\chi^2$, or more physical fit parameters), it
remains difficult to draw a firm conclusion.

To overcome these difficulties, Ref.~\cite{Ducloue:2019ezk} proposed a reorganisation of 
perturbation theory in which the evolution time is the physical rapidity $\eta$. This program
lead to a new version of ``collinearly-improved BK equation'', shown below in~\eqn{bketa}. This equation is non-local in rapidity. In that sense it looks formally similar
to the equation proposed in~\cite{Beuf:2014uia} but these two
equations are fundamentally
different: \texttt{(i)} the evolution variable, $\eta$, occurring in~\eqn{bketa}
is the physical rapidity $\ln(1/x)$, and not the rapidity of the dipole projectile; \texttt{(ii)} the rapidity
shift in~\eqn{bketa} accounts for an all-order resummation of double {\em collinear} logarithms,
and not {\em anti}--collinear (see Sect.~\ref{sect:evol}). The collinear
logarithms are generated by gluon emissions
with relatively large momenta. Such emissions are atypical in the
context of DIS. Moreover, they are partially suppressed by
non-linear effects, so their resummation is somewhat less
critical. As a consequence,
a comparison between various resummation prescriptions shows
only little scheme dependence~\cite{Ducloue:2019ezk}, at the level of the expected perturbative accuracy of the resummed
equation.

The evolution equation we use in this work, \eqn{bketa},
actually goes beyond the original proposal from
Ref.~\cite{Ducloue:2019ezk} by additionally including a class of
DGLAP-like single transverse logarithms (both collinear and
anti-collinear) which appears in the NLO BK equation.
While in principle one can also include~\cite{Ducloue:2019ezk} the full set of NLO BK
corrections in our evolution equation, this is numerically cumbersome
(see however~\cite{Lappi:2016fmu}) and it goes beyond the scope of
this Letter.
Instead, we directly confront the relatively simple equation
\eqref{bketa} with the most recent HERA data~\cite{Abramowicz:2015mha}
for inclusive DIS at small Bjorken $x\le 0.01$. This dataset includes more
data points as compared to the one used in previous
fits~\cite{Iancu:2015joa,Albacete:2015xza}.

We use the fits to test the resummation of the aforementioned double
logs and single transverse logarithms as well as various prescriptions
for the running of the QCD coupling which enters the BK equation. We
study two models for the initial condition to the BK equation: the
simple Golec-Biernat and W\"usthoff (GBW) model
\cite{GolecBiernat:1998js,GolecBiernat:1999qd} and a a
running-coupling version of the McLerran-Venugopalan (MV)
model~\cite{McLerran:1993ka}. Both models include two free parameters
which are fitted to the experimental data. Two additional parameters
(leaving aside the quark masses that we keep fixed) are associated
with our {\it ansatz} for the running coupling and with the overall
normalisation of the DIS cross-section.

With this setup, we find a good overall agreement with the HERA data
at small $x$.
Furthermore, it appears that the physically-motivated MV model with
running coupling is preferred over the GBW model, with the former
giving both better $\chi^2$ and more reasonable parameter values than
the latter.
As with earlier studies, the fits appear to favour an initial condition for the dipole scattering amplitude
with a fast and abrupt approach to the unitarity limit. This is the main reason
why better fits are obtained with the running coupling version of the MV model as compared to
its original fixed-coupling version~\cite{McLerran:1993ka}. This finding is also in agreement
with the fact that previous
fits~\cite{Albacete:2009fh,Albacete:2010sy} using the original MV
model favoured a modified
dependence of the initial amplitude on the dipole size, decreasing
like $r^{2\gamma_0}$ with $\gamma_0>1$.
Another interesting observation of our fits is that the inclusion of
the DGLAP-like single logarithms significantly improves the
description of the data at large $Q^2$, allowing for good descriptions
up to maximum $Q^2$ of 400~GeV$^2$.

This Letter is organised as follows. Sect.~\ref{sect:evol} provides a
qualitative and (hopefully) pedagogical summary of the arguments
justifying the replacement of the original collinear-improved BK
equation formulated in terms of the rapidity of the dipole
projectile~\cite{Beuf:2014uia,Iancu:2015vea}, by a new version
formulated in terms of the rapidity of the hadron target, or Bjorken
$x$. (We refer to Ref.~\cite{Ducloue:2019ezk} for more details.) 
Sect.~\ref{sec:illustrations} discusses the main physical consequences of
the various resummations on the solution to the collinearly-improved BK
equation. Finally, Sect.~\ref{sect:fits}
presents the main original results of this paper: the setup and the 
results for the fits together with their physical discussion.

\section{Collinearly-improved BK evolution in the target rapidity}
\label{sect:evol}

The main difference between our present approach and
previous saturation fits to DIS refers to our specific choice of
evolution equation used to describe the evolution of the dipole
$S$-matrix with increasing energy.
More precisely, we use a collinearly-improved 
version of the BK equation --- recently proposed
in~\cite{Ducloue:2019ezk} --- in which
the rapidity variable playing the role of the evolution time is the {\it proton} rapidity $\eta=\ln(1/x)$,
with $x$ the standard Bjorken variable. This contrasts previous
studies (see
e.g.~\cite{Balitsky:2008zza,Beuf:2014uia,Iancu:2015vea,Iancu:2015joa})
where the evolution was formulated in terms of the rapidity $Y$ of the
dipole {\em projectile}.
Beyond leading order~\cite{Balitsky:1995ub,Kovchegov:1999yj}, the
choice of $\eta$ over $Y$ has important consequences.
To make this clear, we first summarise the main findings
of~\cite{Ducloue:2019ezk} which are relevant for the fit to DIS data
described in the next section.

\paragraph{Basic kinematics, target and projectile rapidity} 
We use a frame in which the virtual photon, $\gamma^*$, is an energetic right-mover
with (light-cone) 4-momentum $q^\mu\equiv (q^+, q^-, \bq_\perp)= (q^+, -\frac{Q^2}{2q^+},
\bm{0}_\perp)$, whereas the proton target is a left-mover with $P^\mu=\delta^{\mu-}P^-$.\footnote{We neglect the
proton mass $M$ which is much smaller than all the other scales in the
problem, $M^2\ll Q^2\ll 2p\cdot q$.}
In the high-energy or small Bjorken $x$ regime, 
\beq\label{xBj}
x\equiv\,\frac{Q^2}{2P\cdot q}=\,\frac{Q^2}{2P^- q^+}\,\ll\,1\,,
\eeq
the coherence time $\Delta x^+\simeq 2q^+/Q^2$ of the virtual photon,
i.e.\ the  typical lifetime of its quark-antiquark ($q\bar q$)
fluctuation, is much larger than the longitudinal extent $1/P^-$ of
the target.
This justifies the use of the dipole picture in which the $\gamma^*$ fluctuates into a $q\bar q$ colour
dipole long before the collision, which then scatters inelastically
off the proton.
Via the optical theorem, the total dipole-hadron scattering
cross-section is related to the $S$-matrix for the elastic
scattering. At high energy, one can work in the eikonal approximation
where the transverse coordinates $\bx$ of the quark and
$\by$ of the antiquark are not affected by the collision.

The elastic $S$-matrix $S_{\bx\by}$ also depends on the rapidity
difference between the dipole and the proton through the {\it
  high-energy evolution}. The physical picture of this evolution and its analytical description
depend on how the total energy is divided between the (dipole)
projectile and the (proton) target i.e. upon the choice of the ``dipole
frame'' in which one is working. It is useful to consider the two
extreme situations: the ``target frame'', in which most of the total
energy (and hence the whole high-energy evolution) is carried by the
proton, and the ``projectile frame'', where the energy is mostly
carried by the dipole (and the high-energy evolution is encoded in the
dipole wavefunction). Importantly the rapidity variable which
represents the ``evolution time'' for the high-energy evolution, is
different in these two situations:
\begin{align}
  \text{target rapidity: }
  & \eta\equiv\,\ln\frac{P^-}{|q^-|}\,=\,\ln\frac{2 q^+P^-}{Q^2}\,=\,\ln\frac{1}{x}\label{eta}\\
  \text{dipole rapidity: }
  & Y\equiv \,\ln\frac{q^+}{q_0^+}=\,\ln\frac{2q^+P^-}{Q_0^2}=\,\ln\frac{1}{x}+
\ln\frac{Q^2}{Q_0^2}=\eta +\rho\label{Ydef}
\end{align}
For the target rapidity, the typical gluon from the proton which
interacts with the dipole has a longitudinal momentum
$k^-={Q^2}/{2q^+}=|q^-|$, and hence a longitudinal extent $\sim 1/k^-$
of the order of the lifetime $\Delta x^+$ of the $q\bar q$ pair.
For the projectile rapidity, the softest dipole to participate in the
collision has a longitudinal momentum $q_0^+=Q_0^2/2P^-$ --- i.e.\ a
lifetime ${2q_0^+}/{Q_0^2}$ equal to the longitudinal extent $1/P^-$
of the proton ---, where the scale $Q_0\ll Q$ is the transverse momentum scale for the
onset of unitarity (multiple scattering) effects in the (unevolved)
proton.
The two rapidities differ by $\rho\equiv \ln({Q^2}/{Q_0^2})$ which is
large when $Q\gg Q_0$. 

The non-linear effects associated with the high gluon density are
described differently in the two frames.
In the target frame, soft gluon emissions occur in the proton
wavefunction which is a dense environment. Accordingly, these
emissions are modified by non-linear effects like gluon
recombinations. The non-linear evolution of the dense hadron
wavefunction has been computed only to leading order, yielding the
(functional) JIMWLK equation
\cite{JalilianMarian:1997jx,JalilianMarian:1997gr,Kovner:2000pt,Iancu:2000hn,Iancu:2001ad,Ferreiro:2001qy}.
Conversely, in the dipole frame one views the evolution as successive,
soft, gluon emissions within the dipole wavefunction, a dilute
hadronic system. Gluon emissions from the dipole occur like in the
vacuum and non-linear effects exclusively refer to multiple
scattering.
This leads to the Balitsky hierarchy (and the BK equation), currently
known to NLO accuracy~\cite{Balitsky:2008zza,Balitsky:2013fea,Kovner:2013ona,Kovner:2014lca,Lublinsky:2016meo}.
Since our purpose in this work is to go beyond LO accuracy, we
systematically use the dipole frame in what follows.

\paragraph{Time ordering and collinear improvements in $Y$ (dipole
  frame)}
Besides being less suited for applications to DIS, the evolution with
$Y$ has a more severe conceptual drawback: the typical emissions contributing to this evolution at leading order
can violate proper time ordering, that is, the condition that the formation time of a daughter gluon
be smaller than the lifetime of its parent.\footnote{Notice that formation times and lifetimes
  are parametrically the same for this space-like evolution.}
To understand this, we first recall that, when
$Q^2\gg Q^2_0$, the typical emissions associated with the high-energy
evolution of the dipole wavefunction are strongly ordered both in
longitudinal momenta ($k^+$) and in transverse momenta ($k_\perp$):
\beq\label{DLAphase}
q^+\gg k_1^+\gg k_2^+\gg\dots\gg q_0^+\,,\qquad Q^2\gg k_{1\perp}^2\gg k_{2\perp}^2
\gg\dots\gg Q_0^2\,.
\eeq
This corresponds to soft and collinear emissions which yield the
dominant, double-logarithmic, contributions proportional to powers of
$\abar Y\rho$. However, an explicit calculation of the relevant
Feynman graphs shows~\cite{Iancu:2015vea}
that this double-logarithmic enhancement only holds so long as the gluon lifetimes 
are ordered as well:
\beq\label{TOcond}
\frac{2q^+}{Q^2}\gg \frac{2k_1^+}{ k_{1\perp}^2}\gg  \frac{2k_2^+}{k_{2\perp}^2}\gg\dots\gg \frac{2q_0^+}{Q_0^2}\,.\eeq
This condition reduces the rapidity phase-space available for the
evolution from $Y$ to $Y-\rho\equiv \eta$.
The condition~\eqn{TOcond} is already violated (due to the emission of
daughter gluons with sufficiently soft $k_\perp$) in the LO BK evolution
which resums an infinite series in $\abar Y\rho$, instead of the correct
series in powers of $\abar(Y-\rho)\rho$.
The difference between the two corresponds to an alternating series of
double ``anti-collinear'' logarithms proportional to $\abar\rho^2$
which spoil the convergence of the perturbative expansion in $Y$.
In particular, the NLO BK equation includes the first (negative)
contribution proportional to $\abar\rho^2$~\cite{Balitsky:2008zza}
which makes the evolution {\it unstable} \cite{Lappi:2015fma} and
hence unsuitable for physical studies.\footnote{More generally there
  is a {\it tower} of series of such spurious terms: series
  appears to correct the time-ordering violation in
  the previous one. The dominant series includes all powers of
  $\abar\rho^2$, the subdominant one, those of $\abar^2\rho^2$, etc.}

To overcome this difficulty, it was originally
proposed~\cite{Beuf:2014uia,Iancu:2015vea} to enforce time-ordering
directly in the dipole frame evolution. Two ``collinearly improved''
BK equations have been proposed: in the first~\cite{Beuf:2014uia} the
evolution is non-local in rapidity and has the same kernel as at LO,
while in the second~\cite{Iancu:2015vea} the evolution is local in
$Y$, but both the kernel and the initial condition receive corrections
to all orders in $\abar\rho^2$.
Both methods allow for a faithful resummation of the dominant series
in powers of $\abar\rho^2$, but the subleading terms (proportional to
powers of $\abar^k\rho^2$ with $k\ge 2$) are not under control.
At a first sight, both strategies appear to be successful:
the respective equations are stable~\cite{Iancu:2015vea,Lappi:2016fmu}, they can be extended to full NLO
accuracy~\cite{Lappi:2016fmu}, and moreover they allow for good
fits to the HERA data for DIS at small $x$~\cite{Iancu:2015joa,Albacete:2015xza}.
 
\paragraph{Recasting dipole evolution in terms of $\eta$}
A more recent study has revealed that these apparent successes
were in fact deceiving~\cite{Ducloue:2019ezk}. The numerical studies
in~\cite{Iancu:2015joa,Albacete:2015xza,Lappi:2016fmu} have been
presented in terms of $Y$ instead of the physical
rapidity $\eta=Y-\rho=\ln(1/x)$, and in the DIS fits in~\cite{Iancu:2015joa,Albacete:2015xza}, 
the variable $Y$ has been abusively interpreted as $\ln(1/x)$.
The correct procedure would require to first transform the results
from $Y$ to $\eta$ by a simple change of variables, before attempting
a physical interpretation or a fit to the data. When following this
correct procedure, one finds \cite{Ducloue:2019ezk} an {\it
  unacceptably large resummation-scheme dependence}. For example when
solved with the same initial condition the two
``collinearly-improved'' BK equations introduced
in~\cite{Beuf:2014uia} and~\cite{Iancu:2015vea} yield very different
predictions for the evolution in $\eta$.\footnote{In particular, they
  predict widely different values for the saturation exponent
  $\lambda_s$ at large $\eta$, see the right panel of Fig.~1
  in~\cite{Iancu:2015vea}, where even the sign of the deviation
  w.r.t.\ the LO value appears to be different in the two schemes.}
More generally, different choices for the ``rapidity shift'' in the
non-local equation in $Y$, albeit equivalent to DLA, lead to very
different predictions for the saturation exponent
$\lambda_s$~\cite{Ducloue:2019ezk}.
This strong scheme dependence forbids any physical interpretation of
the results.  It demonstrates that the uncontrolled, subleading,
double-logarithmic corrections --- which generally differ from one
resummation scheme to another --- are numerically important.

The problem is further complicated by the fact that the resummed BK evolution in $Y$ cannot
be formulated as a genuine initial-value problem. The non-local equation introduced in~\cite{Beuf:2014uia}
must be solved as a boundary-value problem (on a line of constant
$Y-\rho$) which seriously complicates even numerical studies of the
equation.
Moreover, even if the local equation with a resummed
kernel~\cite{Iancu:2015vea} does admit an initial-value formulation,
the corresponding initial condition must itself be resummed to account
for double-logarithmic corrections to all orders, a task which appears
to be intractable beyond strict DLA.

These difficulties with the 
evolution in $Y$ can be avoided altogether by using $\eta$ as an ``evolution time''~\cite{Ducloue:2019ezk}.
This choice has some obvious virtues in practice: $\eta=\ln(1/x)$ is the right variable to be used in
phenomenological studies of DIS and, clearly, a boundary-value condition formulated at constant
$Y-\rho$ becomes an initial condition for the evolution in
$\eta$. Most importantly, one can show that the evolution
in $\eta$ naturally ensures the proper time ordering of the successive
emissions.\footnote{In a nutshell, for a gluon of (projectile)
  rapidity $Y_k=\ln(k^+/q_0^+)$ and transverse momentum $k_\perp$, one
  has $\eta_k\equiv Y_k-\rho_k= \ln\tfrac{\tau_k}{\tau_0}$ with
  $\rho_k=\ln(k_\perp^2/Q_0^2)$ and $\tau_k=2k^+/\kt^2$ is the gluon
  lifetime. Ordering in $\eta$ therefore coincides with ordering in
  proper time.}

Instead of computing directly the target evolution in $\eta$ which would be delicate
in the presence of saturation, Ref.~\cite{Ducloue:2019ezk}
proposed to reformulate the dipole evolution in $Y$,
computed in pQCD, as an evolution in $\eta$ via the change of
variables $Y\equiv \eta+\rho$.
Such a change of variables is unambiguous in perturbation theory and has
been used~\cite{Ducloue:2019ezk} to obtain the NLO BK equation in $\eta$ from the respective
equation in $Y$~\cite{Balitsky:2008zza}.

\paragraph{Resummation of atypical collinear double logarithms}
When using $\eta$ as an evolution variable, the BK equation is not
affected by the large anti-collinear logarithms that were present in
the evolution in $Y$.
This is a consequence of time ordering, \eqn{TOcond}, being automatically satisfied for
the evolution in $\eta$.
Furthermore, \eqref{TOcond} also guarantees that {\em typical}
anti-collinear emissions --- i.e.\ those strongly ordered in
transverse momentum according to~\eqref{DLAphase} --- automatically
satisfy the proper ordering in longitudinal momentum $k^+$
(cf.~\eqref{DLAphase} again).

However, even with the proper time-ordering, the correct ordering in
$k^+$ can still be violated by a series of {\it collinear} emissions
with a strong transverse momentum ordering opposite to that
of~\eqn{DLAphase}.
These violations yield double-logarithmic corrections to the BK
kernel, starting at NLO (cf.~\cite{Ducloue:2019ezk}).
In principle, this problem is as severe as the one of time-ordering
violations in the evolution in $Y$: these collinear logarithms have to
be resummed to all orders in the evolution in $\eta$, as the
anti-collinear were resummed in the evolution in $Y$.
However, these collinear emissions, where the daughter gluon has a
much larger transverse momentum than the parent one are {\em atypical}
in the context of DIS. One can further show that they are also strongly suppressed by
saturation which freezes the evolution for emissions with sufficiently soft
transverse momentum.

The NLO evolution in $\eta$ nonetheless has a contribution from
collinear logarithms which, albeit suppressed, eventually translates
into an instability at large-enough rapidity. In practice one would
therefore resum it to all orders (see~\cite{Ducloue:2019ezk}) using a
rapidity shift leading to a non-local evolution in $\eta$ (see
\eqn{bketa} below). As for the resummations in $Y$, this resummation
scheme is not unique beyond DLA. But unlike what happens with the
resummations in $Y$, the scheme dependence for the resummations in
$\eta$ is reasonably small, in agreement with the expected
perturbative accuracy of the resummed equations. For example, choosing
different resummation schemes (e.g.\ by varying the $\eta$ shift in
\eqref{bketa}), one finds a small impact on the saturation exponents,
comparable with missing perturbative contributions of
${\cal {O}}(\abar^2)$.

\paragraph{The collinearly-improved BK equation in $\eta$}
We are finally in a position to present the evolution equation in the
target rapidity, $\eta$ which reads
\begin{equation}
	\label{bketa}
	\frac{\del {S}_{\bx\by}(\eta)}{\del \eta}  = 
	\int \dif^2\bz \frac{\abar(r_{\min})}{2\pi} 
	\frac{(\bx-\by)^2}{(\bx-\bz)^2(\bz-\by)^2}\,\left[\frac{r^2}{\bar z^2}\right]^{\pm A_1}
	\big[{S}_{\bx\bz}(\eta \minus \delta_{\bx\bz;r})
	{S}_{\bz\by}(\eta \minus \delta_{\bz\by;r}) - {S}_{\bx\by}(\eta) \big], 
\end{equation}
where $\bz$ is the transverse position of the gluon emitted by either
the quark or the antiquark. In the large-$N_c$ approximation, implicit
in~(\ref{bketa}), this
can be viewed as the splitting of the original dipole $(\bx,\,\by)$
into two daughter dipoles $(\bx,\,\bz)$ and $(\bz,\,\by)$.
The other notations are explained below.
Compared to the LO BK equation (in $\eta$) a few key differences are
worth noting:

\texttt{(i)} the use of the one-loop running coupling $\abar(r_{\min})$ with the running scale set by the size
$r_{\min}$ of the smallest dipole: $r_{\min} \equiv
\min\{|\bx-\by|,|\bx-\bz|,|\bz-\by|\}$. Alternative schemes are
discussed in the next section.

\texttt{(ii)} the rapidity shifts in the arguments of the $S$-matrices
for the daughter dipoles ensure the resummation of the leading double logarithms associated with the
$k^+$ ordering. They are given by
\begin{equation}\label{delta}
  \delta_{\bx\bz;r} \equiv \max \left\{0,\ln\frac{r^2}{|\bx\minus\bz|^2} \right\} 
\end{equation}
with $r\equiv|\bx-\by|$, and similarly for $\delta_{\bz\by;r}$. They
are non-zero only for emissions in which one of the daughter dipoles
is (much) smaller than the parent one, in agreement with our earlier
discussion about collinear logarithms.
Expanding~(\ref{bketa}) to first non-trivial order in $\delta$ would
give the double-logarithmic contribution to the BK kernel at NLO which
eventually yields an instability unless properly resummed as
in~(\ref{bketa}).

\texttt{(iii)} \eqn{bketa} also includes the resummation of the first set of single DGLAP logarithms (either collinear, or anti-collinear),
 via the factor $\left[{r^2}/{\bar z^2}\right]^{\pm A_1}$, where $\bar z \equiv \min\{|\bx-\bz|,|\bz-\by|\}$.
The number $A_1=11/12$ is related to the DGLAP splitting function via
the following Mellin transform:
\begin{equation}
   \label{pomega}
	\int_0^1 \dif z\, z^\omega 
	\left[ P_{\rm gg}(z) + \frac{\CF}{\Nc}\, P_{\rm qg}(z) \right]
	= \frac{1}{\omega} - A_1
	+\mathcal{O}\left(\omega,\frac{\Nf}{\Nc^3}\right).
\end{equation} 
The singular piece $1/\omega$ generates the
$\eta=\ln(1/x)$ logarithm contributing to the LO BK evolution, while
the non-singular piece $(-A_1)$ contributes at NLO and is enhanced by
a single transverse logarithm $\ln({r^2}/{\bar z^2})$. The sign in the
exponent, $\pm A_1$, is taken to be plus for an anti-collinear
emission (${r^2}<{\bar z^2}$) and minus for a collinear one, so this
factor is always suppressing the evolution.

\eqn{bketa} has to be solved as an initial value problem, albeit a
somehow unusual one due to its non-locality in $\eta$. Indeed, since
the shift introduces a dependence to rapidities smaller than $\eta$,
if we want to start the evolution at some rapidity $\eta_0$ we should
specify the initial condition for $\eta\le \eta_0$. Our prescription
is to assume a constant behaviour in $\eta$ (see Sect.~9 of
Ref.~\cite{Ducloue:2019ezk}) i.e.
\beq\label{ICeta}
{S}_{\bx\by}(\eta <\eta_0) ={S}^{(0)}_{\bx\by}\,.
\eeq
With this prescription, \eqn{bketa} can be solved and used for DIS fits.

A final note concerns the perturbative accuracy of \eqn{bketa}.
We have argued that it includes all the NLO corrections
enhanced by at least one transverse logarithm. Hence, from the
viewpoint of a strict weak-coupling expansion, it is accurate up to
pure NLO corrections, of $\order{\abar^2}$ without any logarithmic
enhancement. Furthermore, the resummation-scheme dependence of
\eqref{bketa} is also coherent with missing ${\cal{O}}(\abar^2)$ terms.
It is possible to extend this equation to full NLO accuracy by adding
the missing pure $\abar^2$ corrections. The resulting equation, which
can be found in Ref.~\cite{Ducloue:2019ezk}, is substantially more
complex and we  postpone its applications to DIS to future work.

\section{Illustrating the impact of running coupling and
  resummation effects}\label{sec:illustrations}

Before turning to the description of inclusive DIS data, it is helpful
to briefly illustrate how the various ingredients in the BK equation,
namely running-coupling (RC) corrections and the resummation of transverse
logarithms, affect the evolution in $\eta$.
For this purpose, we choose a homogeneous target, i.e.\ take
${S}_{\bx\by}(\eta)=S(\eta,r)$ with $r=|\bx-\by|$, 
together with the simple Golec-Biernat--W\"usthoff (GBW) initial
condition~\cite{GolecBiernat:1998js,GolecBiernat:1999qd}: $
S_0(r)=\exp(-r^2 Q_0^2)$ with $Q_0^2=1\,{\rm GeV}^2$.
This Gaussian {\it Ansatz} does not capture the proper behaviour for
$r^2 Q_0^2\ll 1$ but is enough for illustrating our points in this section.

Let us first discuss running-coupling effects. For the sake of the
present discussion, we use
\begin{equation}\label{alpha0}
\abar(r) = \frac{1}{\bar b_0\ln\big[4/(r^2\Lambda^2)\big]},
\end{equation}
with $\bar b_0=0.75$ (corresponding to $n_f=3$) and
$\Lambda=0.2$~GeV. The Landau pole is avoided by freezing the
coupling at the value $\bar\alpha_{\rm sat}=0.67$.
There is some freedom in implementing RC corrections in the BK
equation and we consider two different prescriptions: the
minimal dipole prescription, $\abar(r_{\min})$, where \eqn{alpha0} is
evaluated at
$r = r_{\min} \equiv \min\{|\bx-\by|,|\bx-\bz|,|\bz-\by|\}$ and the
BLM prescription (also dubbed as
``fast apparent convergence'' \cite{Iancu:2015joa}), defined as
\begin{equation}\label{ablm}
  \bar{\alpha}_{\rm \scriptscriptstyle BLM} = 
  \left[ 
    \frac{1}{\abar(|\bx \minus\by|)}
    + \frac{(\bx\minus\bz)^2 - (\bz\minus\by)^2}{(\bx\minus\by)^2}
    \frac{\abar(|\bx-\bz|)-\abar(|\bz-\by|)}{\abar(|\bx-\bz|)\abar(|\bz-\by|)}
  \right]^{-1}.
\end{equation}
Other prescriptions, not studied here, are also possible (see e.g. 
\cite{Balitsky:2006wa,Kovchegov:2006vj,Balitsky:2008zza}). They all have
in common that they reduce to $\abar(r_{\min})$
when one of the three dipoles is much smaller than the other two, as
one can easily check for $\bar{\alpha}_{\rm \scriptscriptstyle BLM}$.
This minimises the NLO correction to the BK equation
proportional to the one-loop $\beta$-function.

Besides varying the prescription for the running coupling, we also aim
to probe the effect of the resummation of transverse logarithms. Single
logarithms can be switched off by removing the factor
$\left[{r^2}/{\bar z^2}\right]^{\pm A_1}$ in \eqn{bketa}, while to
remove double transverse logarithms we set the $\eta$ shifts,
$\delta_{\bx\bz;r}$ and $\delta_{\bz\by;r}$, to zero.

\begin{figure}
  \centerline{%
    \includegraphics[width=0.33\textwidth,page=1]{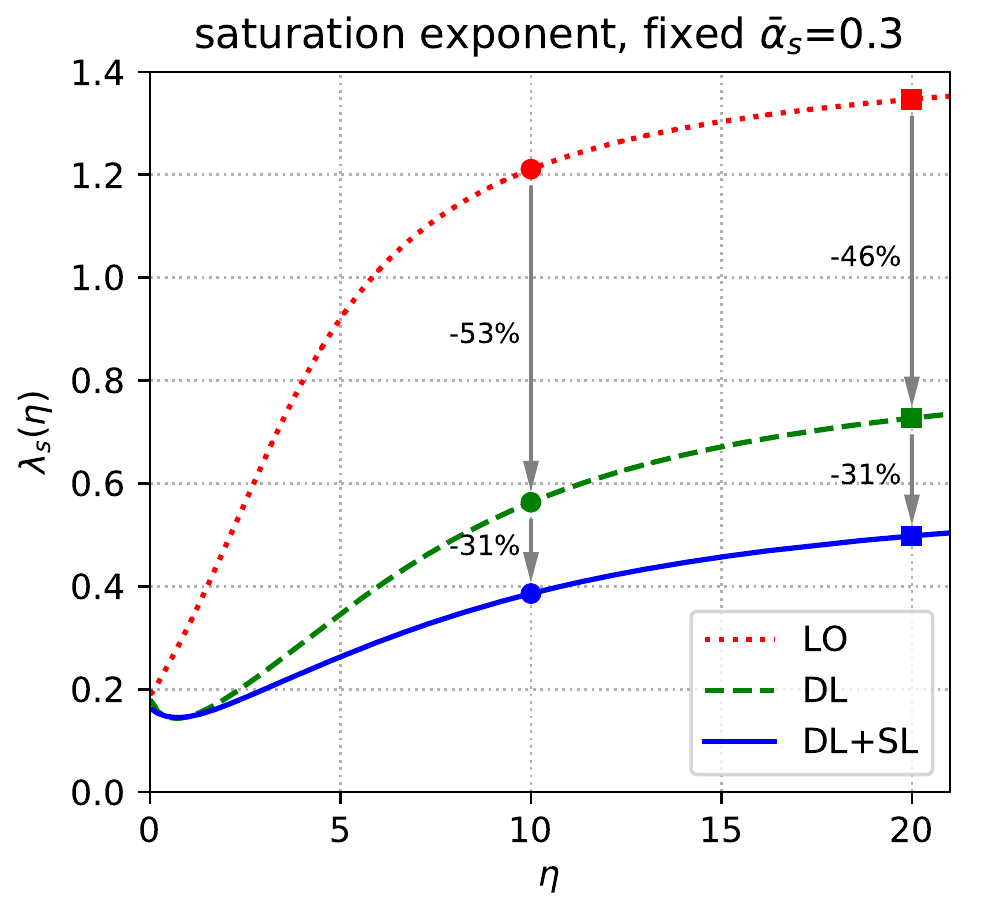}\hfill%
    \includegraphics[width=0.33\textwidth,page=2]{lambda.pdf}\hfill%
    \includegraphics[width=0.33\textwidth,page=3]{lambda.pdf}%
  }
  \caption{The saturation exponent $\lambda_s(\eta)$ extracted from
    solutions to \eqn{bketa} with GBW initial condition and with 3
    different choices for the QCD coupling: fixed coupling $\abar=0.3$
    (left), RC $\abar(r_{\min})$ (middle), and RC
    $\bar{\alpha}_{\rm \scriptscriptstyle BLM}$ (right). In each of
    these 3 cases, we show results corresponding to LO, LO+DL, and
    LO+DL+SL, respectively (see text for details).}
  \label{fig:lambdas}
\end{figure} 

In practice, we solve \eqn{bketa} numerically up to $\eta=20$ and
study the saturation exponent defined as
\beq
\lambda_s(\eta)\,\equiv\,\frac{\rmd\ln Q_s^2(\eta)}{\rmd\eta}\,,
\eeq
with the saturation momentum $Q_s(\eta)$ numerically obtained from the
condition that $S(\eta,r)=\tfrac{1}{2}$ when $r=2/Q_s(\eta)$.
The rapidity range under study is not sufficient to reach the
universal asymptotic behaviour but is representative for the actual range
covered by the HERA data, $\eta\lesssim 10$.

Our results are presented in Fig.~\ref{fig:lambdas} for a
fixed-coupling prescription as well as for our two RC
prescriptions. In each case we show the saturation exponent for the LO
BK equation without any transverse logarithm resummation (``LO'', dotted
red), when double transverse logarithms are resummed (``DL'', dashed
green) and when both double and single transverse logarithms are
resummed (``DL+SL'', solid blue). 

The first observation is that running coupling effects have a large
impact, reducing the saturation exponent by $\sim$75\% for both RC
prescriptions at $\eta=10$ and even more at larger rapidities.
The effect of resumming large transverse logarithms is smaller but
still clearly visible: with RC, we see an additional $\sim 10-25\%$
reduction coming from the resummation of double logarithms and a
$\sim 10-20\%$ reduction coming from single logarithms. 
(The  effect is considerably larger when using a fixed coupling.)
The fact that single-log effects are almost as large as double-log
effects is most likely due to the fact that while the latter are only
relevant for large dipole sizes --- where they are reduced by
saturation effects --- the former have an impact on both large
(collinear) and small (anti-collinear) dipoles.

Furthermore, one sees that, without the resummation of the transverse
logarithms, $\lambda_s$ is still
quite large (e.g.\ $\lambda_s(\eta=10)=0.26$ for rcBK with the BLM
prescription).
This seems to be still too large to optimally accommodate the
small-$x$ HERA data. It also likely explains why previous fits based
on
rcBK~\cite{Albacete:2009fh,Albacete:2010sy,Kuokkanen:2011je,Lappi:2013zma}
appeared to prefer another RC prescription, due to Balitsky
\cite{Balitsky:2006wa}, which predicts smaller values for
$\abar$.

\begin{figure}
  \centerline{%
    \includegraphics[width=0.33\textwidth,page=1]{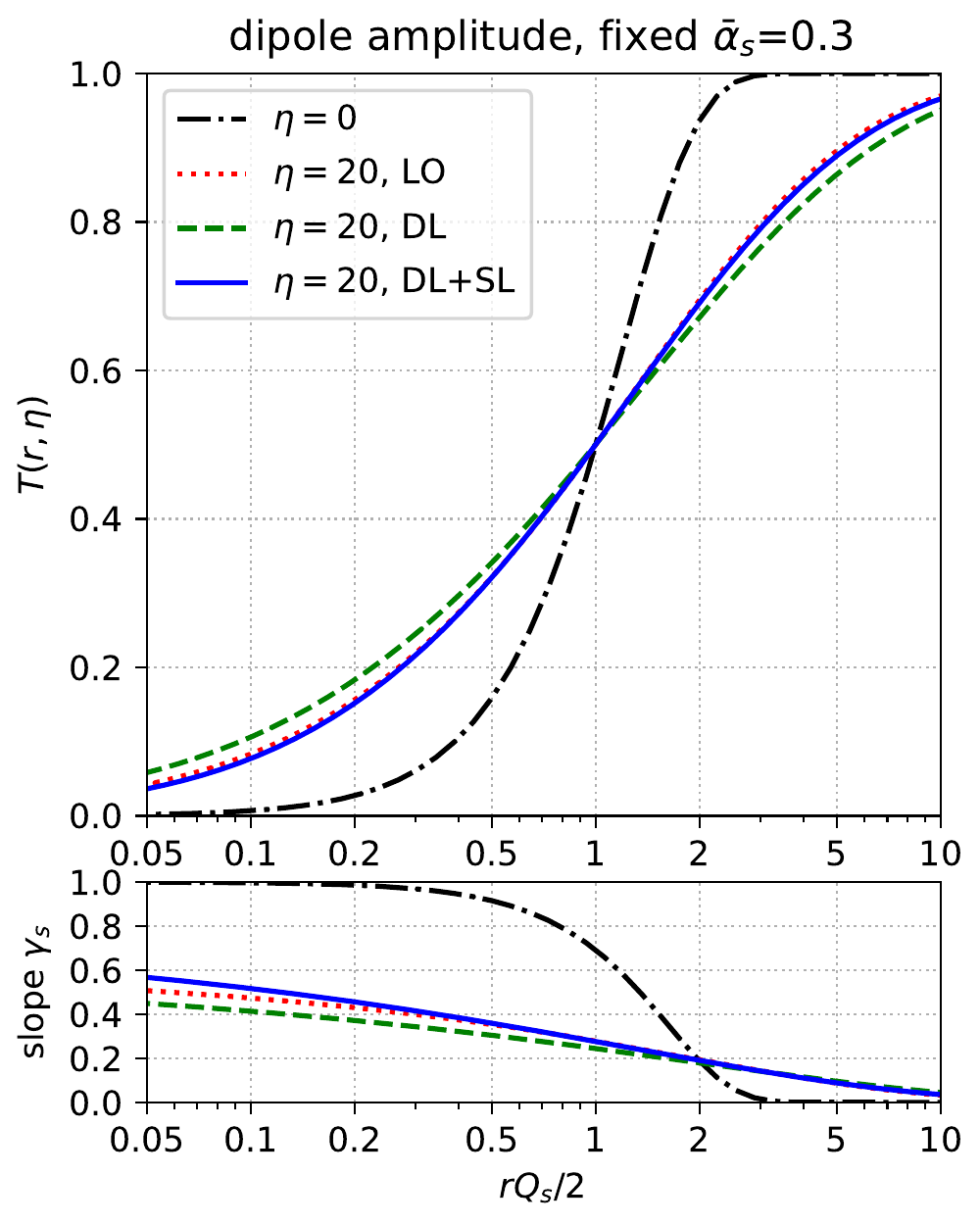}\hfill%
    \includegraphics[width=0.33\textwidth,page=2]{plot-slopes.pdf}\hfill%
    \includegraphics[width=0.33\textwidth,page=3]{plot-slopes.pdf}%
  }
  \caption{Top row: the dipole amplitude plotted as a function of
    the dimensionless quantity $rQ_s/2$. Bottom row: corresponding
    effective slope $\gamma_s=\del\ln(T)/\del\ln(r^2)$. The 3 plots
    correspond to the same configurations as in Fig.~\ref{fig:lambdas}.}
  \label{fig:gammas}
\end{figure}

Beside the saturation exponent which characterises the ``speed'' of
the evolution, it is interesting to look at the form of the amplitude
itself. Both the initial condition (dash-dotted black) and the
amplitude at $\eta=20$ for different degrees of resummation (cf.\
Fig.~\ref{fig:lambdas}) are shown in Fig.~\ref{fig:gammas}, for the
same three configurations as in Fig.~\ref{fig:lambdas}. Results are
plotted as a function of the dimensionless ratio $rQ_s/2$ so as to
better highlight the shape of the dipole amplitude.
The bottom part of the plot shows the effective slope
\begin{equation}\label{eq:gammas}
\gamma_s(r,\eta) = \frac{\del\ln(T(r,\eta))}{\del \ln(r^2)}.
\end{equation}

Two striking features can be observed from these plots.
First, the resummations of double and single transverse logarithms have
a very small impact on the form of the amplitude. Even though a small
effect is visible for the fixed-coupling evolution, almost no effect
is observed with either running-coupling prescriptions.
Then, one sees that the evolved amplitude has a less sharp transition
between the dilute (small-$r$) and saturation (large-$r$) regions.
This is particularly visible for the effective slope $\gamma_s$ which
grows very slowly towards 1 at small dipole sizes.

\section{Fits to the HERA data for inclusive DIS}
\label{sect:fits}

We now come to our main task which is to describe the HERA data~\cite{Abramowicz:2015mha}  for the inclusive DIS cross-section at $x\le 0.01$.

\subsection{The dipole factorisation and the fit set-up}

The dipole factorisation for DIS at small $x$ (and at leading order in pQCD) 
expresses the physical picture in which the virtual
photon fluctuates into a $q\bar q$
pair with a lifetime much longer than the longitudinal extent of the
proton target. The total $\gamma^*p$ cross-section therefore
factorises as a wavefunction for the $\gamma^*\to q\bar q$ splitting
and an interaction between the dipole and the proton:
\begin{equation}
\label{sigmalt}
	\sigma^{\gamma^*p}_{\rmL, \rmT} (x,Q^2)= 
	\sum_f \int \dif^2\br
	\int_0^1 \dif z
	\big| \Psi_{\rmL,\rmT}^{(f)}(r,z;Q^2) \big|^2\,\sigma_{\rm dipole}(\eta_f, r),
\end{equation}
where the squared light-cone wavefunctions (below, $\bar{Q}_f^2 \equiv z(1-z)Q^2 + m_f^2$)
\begin{align}
\label{psil}
&\big| \Psi_{\rmL}^{(f)}(r,z;Q^2) \big|^2 = 
e_q^2 \frac{\alpha_{\rm em} N_c}{2\pi^2}\,
4 Q^2 z^2 (1-z)^2
\rmK_0^2\big(r \bar{Q}_f\big),
	\\
\label{psit}	
&\big| \Psi_{\rmT}^{(f)}(r,z;Q^2) \big|^2 =	
e_q^2 \frac{\alpha_{\rm em} N_c}{2\pi^2}\,
\left\{ \bar{Q}_f^2 \left[z^2+(1-z)^2\right] 
\rmK_1^2\big(r \bar{Q}_f\big) + m_f^2 \rmK_0^2\big(r \bar{Q}_f\big) 
\right\},
\end{align}
represent the probability that a (longitudinal or transverse) virtual
photon splits into a $q\bar q$ colour dipole with transverse size $r$
and with ``plus'' longitudinal momentum fractions $z$ and $1-z$ for
the quark and antiquark respectively. The sum in Eq.~\eqref{sigmalt}
runs over the quark flavors and our fit includes contributions from
the three light quarks with $m_{u,d,s} = 100$~MeV and from the charm
quark, with $m_{c} = 1.3$~GeV.\footnote{We have checked that the
  quality of the fit remains similar for other mass values.}
Furthermore, $\sigma_{\rm dipole}(\eta_f, r)$ is the dipole-proton
total cross-section, evaluated for a proton (target) rapidity
\beq
\eta_f\equiv\ln\frac{1}{\tilde{x}_f}\quad\qquad\mbox{with}\qquad \tilde{x}_f\equiv x \big( 1 + 4 m_f^2/Q^2 \big)\,.
\eeq
In principle this cross-section should be computed by integrating the dipole scattering amplitude $T=1-S$ over
all impact parameters, 
but since the impact-parameter dependence is non-perturbative and not properly encoded in
the BK equation, we simply assume, in the spirit of a mean field picture, that the proton is a uniform
disk of radius $R_p$ (treated as a fit parameter): $\sigma_{\rm dipole}(\eta_f, r)=2 \pi R_p^2 T(\eta_f, r)$,
with $T(\eta, r)$ obtained from numerical solutions to the homogeneous version of  \eqn{bketa}.

One peculiar feature about Eq.~\eqref{sigmalt} is the fact that the
dipole cross-section is evaluated at the rapidity scale
$\eta_f=\ln(1/\tilde{x}_f)$, which refers to the virtual photon, and
not to the dipole.
This looks natural from the perspective of the {\it target} evolution in $k^-$ (recall the discussion
following \eqn{xBj}), but it might look less obvious when thinking about the evolution of the
{\it dipole}. Clearly, the dipole and the virtual photon have different rapidities, due to the splitting
fraction $z$, which can be very asymmetric, i.e. $z\ll 1$ or $1-z\ll 1$, especially  for 
 a transverse photon .
We show however in~\ref{sec:app} that changes in the longitudinal and the
transverse phase-spaces associated with the $\gamma^*\to q\bar q$
splitting compensate each other and that the use of the photon
rapidity $\eta_f$ in~\eqref{sigmalt} is valid.

The quantity we actually fit is the reduced cross-section, related to
the $\gamma^*p$ cross-sections\eqref{sigmalt} by
 \begin{align}
 \label{redfl}
  \sigma_{\rm red} & = \frac{Q^2}{4\pi^2\alpha_{\rm em}}
 \left[\sigma_{\rm T}^{\gamma^*p}+
 \frac{2(1-y)}{1+(1-y)^2}\sigma_{\rm L}^{\gamma^*p}\right].
 \end{align}
$y$ is the inelasticity parameter defined through $Q^2=xys$, with 
$s$ the squared centre-of-mass energy of the $ep$ collision.

In order to solve the evolution equation \eqref{bketa}, we still need to specify the running coupling prescription
and the initial condition. For the running of the QCD coupling, we use the one-loop expression
\begin{equation}
\alpha_s(k_\perp^2) =
\frac{1}{b_{\Nf}\ln\big(k_\perp^2/\Lambda_{\Nf}^2\big)}
\qquad \text{with } \qquad b_{\Nf}=\frac{11 N_c - 2 \Nf}{12 \pi},
\end{equation}
where the dependence on the number of active flavours $N_f$ is made explicit. The value of $\Lambda_{5}$ is determined by imposing $\alpha_s(M_Z^2)=0.1181$~\cite{Tanabashi:2018oca}, and $\Lambda_{3,4}$ are fixed by the continuity of $\alpha_s$ at the flavour
thresholds. We use $m_c=1.3$~GeV and $m_b=4.5$~GeV for the charm and bottom quark masses. In coordinate space we use
\begin{equation}
\label{alphafit}
\alpha_s(r) =
\frac{1}{b_{\Nf}\ln\big[4C_\alpha^2/(r^2\Lambda_{\Nf}^2)\big]},
\end{equation}
where we included a fudge factor $C_\alpha$, which will be one of the
parameters to be fitted to the data. Finally, we freeze $\alpha_s$ at
a value $\alpha_{\rm sat}=1$ to regularise its infrared behaviour.

For the initial condition of the BK evolution, we use two parametrisations, most
conveniently written for the dipole scattering amplitude
$T(\eta,r)\equiv 1-S(\eta,r)$.  The first choice corresponds to a
modified version of the Golec-Biernat and W\"{u}sthoff (GBW)
``saturation model''~\cite{GolecBiernat:1998js} (used already in Fig.~\ref{fig:lambdas}):
\begin{equation}
\label{eq:gbwinit}
{T}(\eta_0,r)= 
\left\{1 -\exp\left[-\left(\frac{r^2Q_0^2}{4}
\right)^p\right]\right\}^{1/p},
\end{equation}
where $Q_0$ and $p$ are parameters to be fitted. The parameter $p$,
not present in the original GBW model, controls the shape of the
amplitude close to saturation. Our second choice is a running-coupling
version of the McLerran-Venugopalan (MV) model~\cite{McLerran:1993ka},
called rcMV in the following, which reads:
\begin{equation}
\label{eq:runinit}
{T}(\eta_0,r)=
\left\{1-\exp\left[-\left(\frac{r^2Q_0^2}{4}\,
\bar\alpha_s(r)\left[1+
\ln\left(\frac{\bar\alpha_{\rm sat}}
{\bar\alpha_s(r)}
\right)\right]\right)^p\right]\right\}^{1/p}.
\end{equation}
Once again, $Q_0$ and $p$ are free parameters (and $\alpha_{\rm sat}=1$). 
The other two parameters of the fit are $C_\alpha$ and $R_p$, the proton radius. 
In practice, we restrict $p$ and $C_\alpha$ to values smaller than 4 and 10, respectively.
Indeed, the natural value for $p$ in the MV model is $p=1$ and it would be reassuring that the fits
prefer such a value. Similarly, a natural value for  $C_\alpha$ should be of order one.

\subsection{Fit results}

\begin{table}[t]
  \begin{center}
    \begin{tabular}{|l|l|l|l|c|c|c|c|c|}
      \hline
      \multicolumn{1}{|c|}{initial} & \multicolumn{1}{|c|}{RC} 
      & \multicolumn{1}{|c|}{double} & \multicolumn{1}{|c|}{single}
      & $\chi^2$ per & \multicolumn{4}{c|}{parameters} \\
      \cline{6-9}
      \multicolumn{1}{|c|}{condition} & \multicolumn{1}{|c|}{scheme}
      & \multicolumn{1}{|c|}{logs} & \multicolumn{1}{|c|}{logs}
      & data point & $R_p$[fm] & $Q_0$[GeV] & $C_\alpha$ &  $p$ \\
      \hline
           &       & no  & no  & 4.33 & 0.671 & 0.414 & 10   & 4 \\
      GBW  & small & yes & no  & 2.05 & 0.786 & 0.355 & 10   & 4 \\
           &       & yes & yes & 1.18 & 0.795 & 0.362 & 5.46 & 4 \\
      \hline
           &       & no  & no  & 1.88 & 0.764 & 0.374 & 10    & 4 \\
      GBW  & BLM   & yes & no  & 1.65 & 0.888 & 0.319 & 6.65  & 4 \\
           &       & yes & yes & 1.14 & 0.762 & 0.377 & 0.788 & 4 \\
      \hline
           &       & no  & no  & 3.89 & 0.655 & 0.659 & 10   & 4 \\
      rcMV & small & yes & no  & 1.72 & 0.757 & 0.569 & 10   & 4 \\
           &       & \bf yes & \bf yes & \bf 1.03 & \bf 0.772 & \bf 0.561 & \bf 5.66 & \bf 1.76 \\
      \hline
           &       & no  & no  & 1.46 & 0.742 & 0.596 & 10    & 4    \\
      rcMV & BLM   & yes & no  & 1.31 & 0.841 & 0.500 & 5.68  & 4    \\
           &       & \bf yes & \bf yes & \bf 1.01 & \bf 0.758 & \bf 0.503 & \bf 0.897 & \bf 1.01 \\
      \hline
    \end{tabular}
  \end{center}
  \caption{$\chi^2$ and values of the parameters fitted to HERA data.}
  \label{tab:fit}
\end{table}

In Table~\ref{tab:fit} we quote the results of the fit to the combined
HERA data for the reduced cross section~\cite{Abramowicz:2015mha}. We
include in the fit data points with $x<0.01$ and $Q^2<50$ GeV$^2$. For
both initial conditions (GBW and rcMV) we show the results
obtained using the two running coupling prescriptions (smallest dipole or
BLM) in three cases corresponding to different
resummations of transverse logarithms: pure LO (rcBK) evolution,
resumming only double logarithms, or resumming both double and single logarithms.

One can see that the resummation of the double and of the single
logarithms are both improving the agreement with the data. This is
particularly the case for the resummation of single transverse
logarithms which not only have a significant impact on the quality
($\chi^2$) of the fit, but also lead to more physical values for the
fit parameters (in particular $C_\alpha$, for which values much larger
than 1 mean that the evolution needs to be artificially slowed down by
an unnaturally small value of the coupling in order to be compatible
with data).\footnote{For the rcMV initial condition and the BLM
  running coupling scheme, discarding double-log corrections but
  including single-log ones even leads to a fit with a $\chi^2$ per
  point of 0.98, with reasonable values of all the fit parameters.}
These two resummations allow to reach values of $\chi^2$ per data
point of less than 1.2 for all the combinations of initial conditions
and running coupling prescriptions considered here.
We believe (see e.g.\ the discussion in
section~\ref{sec:illustrations}) that this good agreement with the
data is largely due to the reduction of the saturation exponent,
$\lambda_s$, due to the resummation of transverse logs.
A particular consequence is that, contrary to some previous rcBK fits,
we do not need to use the peculiar Balitsky prescription for the
running coupling~\cite{Balitsky:2006wa} to obtain a good agreement
with the data. One notices also that the fit shows a slight
preference for the BLM running-coupling scheme which predicts slightly
smaller saturation exponents compared to the smallest dipole prescription.

\begin{figure}
	\centerline{\includegraphics[height=6.5cm]{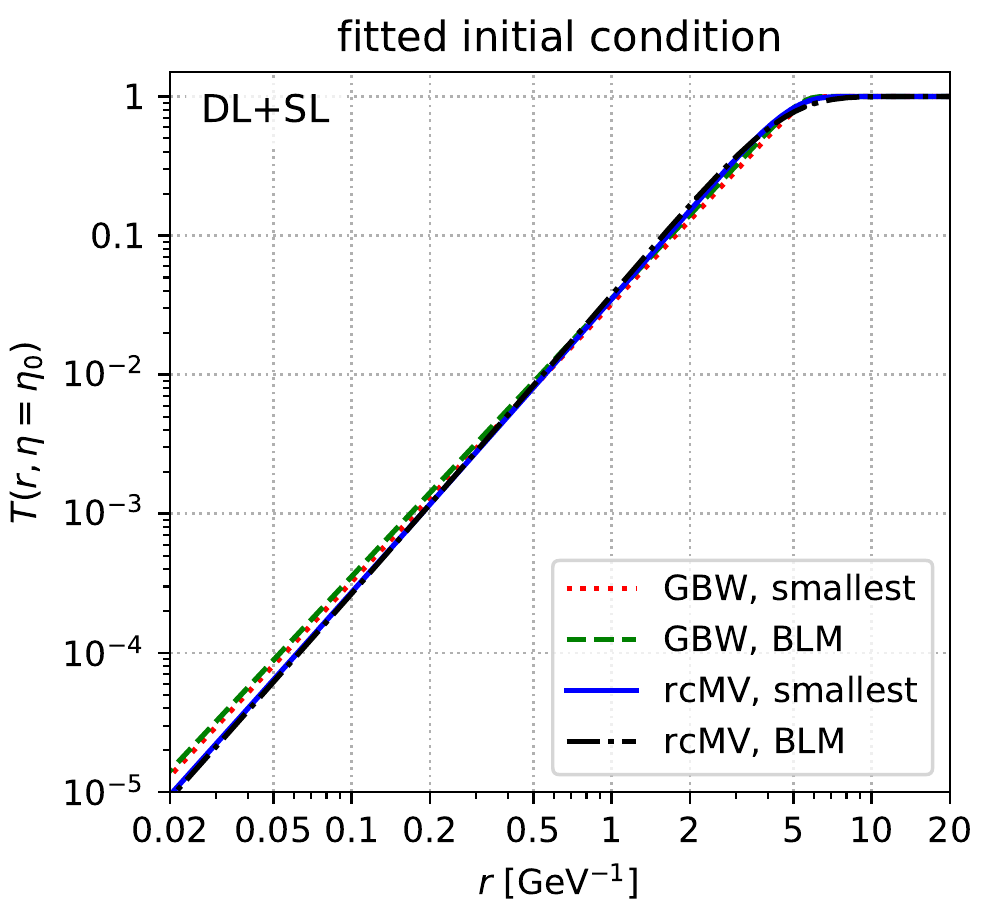}
	\hfill\includegraphics[height=6.5cm]{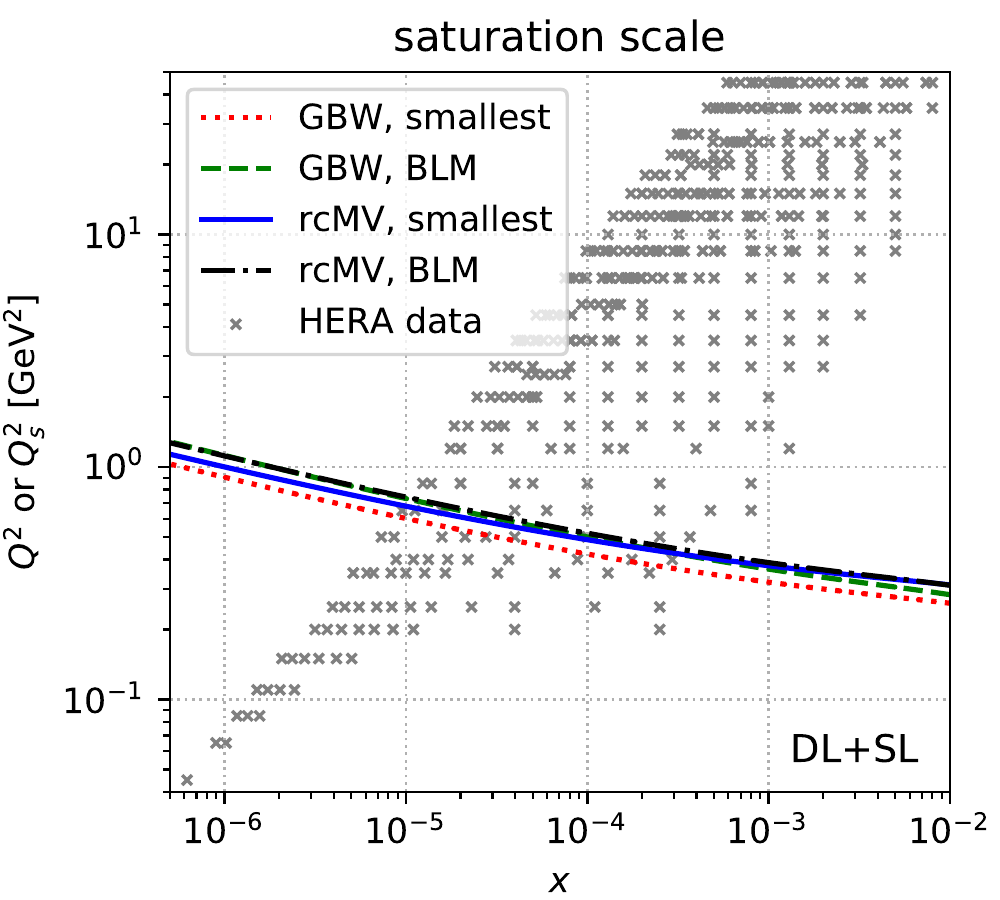}}
	\caption{Left: initial conditions obtained from the fit as described in the text. Right: value of the saturation momentum, defined as $T(r=2/Q_s(\eta), \eta)=1/2$. The experimental data points from HERA~\cite{Abramowicz:2015mha} are shown for comparison.}
  \label{fig:Qs}
\end{figure}

In Fig.~\ref{fig:Qs}~(left) we show the shape of the initial condition as a function of $r$ for the four fits which take into account the resummation of both single and double logarithms. The results look quite similar despite the different functional forms, which is due to the rather strong constraints from the data. In the right panel of Fig.~\ref{fig:Qs}, we show the saturation scale as a function of $x$ for each of these four fits, superimposed over the HERA data points that we use in the fit.

In Table~\ref{tab:largeQ2} we show how the fit is affected by
including data with $Q^2$ larger than 50 GeV$^2$. {\it A priori}, one
would not expect a very good agreement with data at such high
transverse scales, where DGLAP effects are expected to be
essential. Nevertheless one can see that, when including the single
logarithms resummation, the fit quality remains about the same up to
$Q^2_\text{max}=400$ GeV$^2$ for a rcMV initial condition. This is
likely due to the fact that, as explained in section~\ref{sect:evol},
these logarithms represent an important subset of the DGLAP
contributions (at least at small $x$) and therefore improve the
description at large $Q^2$. Even when resumming the single transverse 
logarithms, the fit quality gets worse when going to larger $Q^2$
with a GBW-type initial condition~(\ref{eq:gbwinit}). This is probably
related to the fact that the GBW model does not have the correct
physical behaviour at large transverse momenta, or small dipole sizes.

\begin{table}
  \begin{center}
    \begin{tabular}{|l|l|l|l|c|c|c|c|}
      \hline
      \multicolumn{1}{|c|}{initial} & \multicolumn{1}{|c|}{RC}
      & \multicolumn{1}{|c|}{double} & \multicolumn{1}{|c|}{single}
      & \multicolumn{4}{c|}{$\chi^2$ per point vs.\ $Q^2_{\rm max}$} \\
      \cline{5-8}
      \multicolumn{1}{|c|}{condition} & \multicolumn{1}{|c|}{scheme}
      & \multicolumn{1}{|c|}{logs} & \multicolumn{1}{|c|}{logs}
      & 50   & 100  & 200  & 400  \\
      \hline
           &       & no   & no   & 4.33 & 4.33 & 4.22 & 4.06 \\
      GBW  & small & yes  & no   & 2.05 & 2.17 & 2.27 & 2.24 \\
           &       & yes  & yes  & 1.18 & 1.21 & 1.31 & 1.39 \\
      \hline
           &       & no   & no   & 1.88 & 1.93 & 2.04 & 2.07 \\
      GBW  & BLM   & yes  & no   & 1.65 & 1.75 & 1.94 & 2.01 \\
           &       & yes  & yes  & 1.14 & 1.17 & 1.25 & 1.32 \\
      \hline
           &       & no   & no   & 3.89 & 4.01 & 3.97 & 3.90 \\
      rcMV & small & yes  & no   & 1.72 & 1.86 & 1.93 & 1.92 \\
           &       & \bf yes  & \bf yes  & \bf 1.03 & \bf 1.04 & \bf 1.01 & \bf 1.00 \\
      \hline
           &       & no   & no   & 1.46 & 1.50 & 1.50 & 1.47 \\
      rcMV & BLM   & yes  & no   & 1.31 & 1.34 & 1.35 & 1.33 \\
           &       & \bf yes  & \bf yes  & \bf 1.01 & \bf 1.03 & \bf 1.01 & \bf 1.00 \\
      \hline
    \end{tabular}
  \end{center}
  \caption{Evolution of the fit quality when increasing $Q^2_{\rm max}$ (in GeV$^2$).}
\label{tab:largeQ2}
\end{table}

Since we take into account the charm contribution to the inclusive reduced cross section $\sigma_\text{red}$ in Eq.~(\ref{redfl}), we could also in principle compute the charm production cross section $\sigma_\text{red}^{c\bar{c}}$. In~\cite{Iancu:2015joa} it was found that, after fitting the data for the inclusive cross-section $\sigma_\text{red}$, a very good $\chi^2$/ndf ($<0.7$) was also obtained --- without further tuning the fit parameters --- for the $\sigma_\text{red}^{c\bar{c}}$ data presented in~\cite{Aaron:2009af}. We believe that this was mostly a coincidence since a comparison between the best fits in \cite{Iancu:2015joa} and 
 the more recent, combined, HERA data for charm
 production~\cite{H1:2018flt}, with more points and smaller
 uncertainties, shows a much worse agreement ($\chi^2/\text{ndf}
 >4$). The situation is similar with the present fit. This should not
 be a surprise as the inclusion of heavy flavour data in the
 saturation fits based on the BK equation is a longstanding issue.
 The resummations considered here are not expected to bring any
 concrete improvement on this issue, as they are not adapted to the inclusion of heavy quarks.

\subsection{Positivity of the dipole amplitude's Fourier transform}
\label{sec:FT}

In this work we used the initial condition parametrisations
(\ref{eq:gbwinit}) and (\ref{eq:runinit}) proposed
in~\cite{Iancu:2015joa}. As was later shown~\cite{Giraud:2016lgg}, a
drawback of these expressions is that their Fourier transforms are not
positive-definite, which can lead to unphysical results in momentum
space such as a negative unintegrated gluon distribution. The original
GBW~\cite{GolecBiernat:1998js} and MV~\cite{McLerran:1993ka}
parametrisations are not affected by this issue, however we were only
able to obtain rather poor fits (with $\chi^2/\text{ndf}>1.4$) when using
these expressions. A similar discussion applies to
the more recent MV$^e$ form, introduced
in~\cite{Lappi:2013zma}, which involves an extra
parameter $e_c$ and reads
\begin{equation}
\label{ic_mve}
T(\eta_0,r)=
1-\exp\left[-\frac{r^2Q_0^2}{4} \ln\left(\frac{1}{r \Lambda}+e_c \cdot e\right)\right].
\end{equation}

\begin{figure}
	\centerline{\includegraphics[height=6.5cm,page=1]{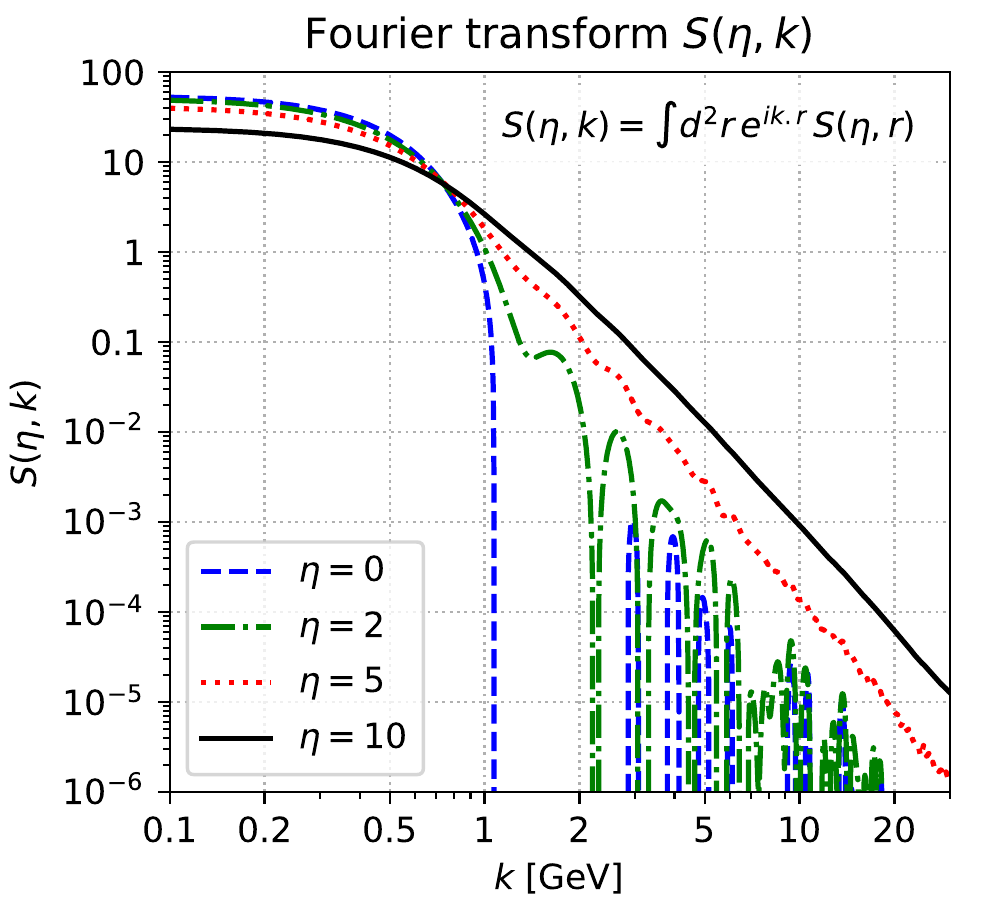}
	\hfill\includegraphics[height=6.5cm,page=2]{SkWk.pdf}}
	\caption{The Fourier transforms $S(\eta,k)\equiv\int \rmd^2\br \,\rme^{\rmi \bk\cdot\br}\, S(\eta,r)$
(left figure) and 	$W(\eta,k)\equiv\int \rmd^2\br \,\rme^{\rmi \bk\cdot\br} \,[T(\eta,r)/{r^2}]$ (right figure), 
as computed from
numerical solutions to \eqn{bketa} with the rcMV initial condition and the BLM prescription for the RC.
(The values of the free parameters are taken from the last line
in Table~\ref{tab:fit}.) The discontinuities in the curves corresponding to the initial condition ($\eta=0$)
and the early rapidities reflect the negativity problem mentioned in Sect.~\ref{sec:FT}.}
  \label{fig:Sk}
\end{figure}

That said, it should be pointed out that the high-energy evolution
tends to improve the situation. If we concentrate on our fits using 
the rcMV initial condition, which is physically most appealing, we numerically
find that, even though the Fourier transform of the initial condition has negativity issues,
the evolved amplitude becomes positive after a few units of
rapidity, cf. Fig.~\ref{fig:Sk} (left). (Some small oscillations remain in an intermediate
  rapidity range before disappearing at larger rapidities.)
This happens because the solution develops an effective slope, defined in Eq.~\eqref{eq:gammas}, which is smaller than 1 even for small $r$ as can be easily seen in the lower row in Fig.~\ref{fig:gammas}. This should be contrasted to the small-$r$ behaviour of the rcMV initial condition for $T(r)$, which vanishes faster than $r^2$ when $r\to 0$.  In fact, after evolution, not only $S(r)$ but also $T(r)/r^2$ (whose Fourier transform is proportional to the gluon occupation number in the proton) satisfy all the conditions given in \cite{Giraud:2016lgg} which are necessary for a function to have a positive Fourier transform.   This is confirmed by the numerical results displayed in Fig.~\ref{fig:Sk}.

Finding a functional form which has a positive-definite Fourier transform while preserving the 
good agreement with the HERA data would be extremely interesting, 
as this could open the way towards a unified description  (via
the dipole factorisation) of inclusive DIS and of particle production in ``dilute-dense'' ($ep$, $eA$, $pp$, $pA$) collisions. However, this seems to be also very challenging.
At this level, it is legitimate to ask whether this difficulty solely reflects our inability to imagine
versatile enough parametrisations for the initial condition, or it rather points towards a deeper problem
with the dipole picture in the presence of a running coupling
(perhaps similar to the problem discussed in \cite{Ducloue:2017dit} in the context of $pA$ collisions).
However, addressing such deep issues goes well beyond the scope of the present work. Our main goal here
was to show that, for a given and physically-motivated initial condition, 
the use of the collinearly-improved version of the BK evolution improves over the standard rcBK
dynamics when it comes to a description of the HERA data for inclusive DIS at small $x$.

\section*{Acknowledgements} 

D.N.T.~would like to acknowledge l'Institut de Physique Th\'eorique de Saclay for hospitality. Part of the work of B.D, E.I. and G.S. has been supported by the Agence Nationale de la Recherche project ANR-16-CE31-0019-01. 
 Part of the work of B.D has been supported by the ERC Starting Grant 715049 ``QCDforfuture''.

\appendix
 
\section{On the rapidity phase-space for dipole evolution}
\label{sec:app}

Eq.~\eqref{sigmalt} uses the rapidity 
argument\footnote{For simplicity, we neglect the quark masses for the sake of this discussion.}
$\eta\equiv \ln(1/x)=\ln(P^-/|q^-|)$
which is the logarithmic phase-space for the proton 
evolution down to the ``minus'' longitudinal momentum
$|q^-|=Q^2/(2q^+)$ of the virtual photon.
This might seem odd since it makes explicit reference to the (longitudinal and
transverse) kinematics of the virtual photon, and not to that of the dipole.
One might expect that the dipole-proton scattering 
amplitude ${T}$ must be evolved over a smaller rapidity interval,
since some of the longitudinal phase space is ``consumed'' to produce
the splitting of the virtual photon into the $q \bar{q}$ pair.
This issue is particularly important for a 
transverse photon, since the  cross-section $\sigma_{\rm T}^{\gamma^*p}$ is 
dominated\footnote{The phase-space for the integral over $r^2$ in the transverse sector
shows  a logarithmic enhancement for such asymmetric configurations in the weak-scattering
regime where $T\propto r^2$.}  (at relatively high $Q^2$)  
by ``aligned jet'' configurations, such that $z$ is either very small, or very close to one.
Moreover, it does not look natural that the dipole amplitude depends on $Q^2$ (via $\eta$); 
instead, it should depend on the dipole size $r^2$. 

We argue here that the simultaneous changes in the longitudinal and the transverse phase-spaces
associated with the $\gamma^*\to q\bar q$ splitting conspire to give a rapidity interval for the evolution
of the dipole amplitude which is still equal to $\ln(1/x)$. We first observe that the  difference $\hat Y$ 
in ``projectile'' rapidity
between the dipole and the proton is fixed by the dipole leg with the smallest
``plus'' longitudinal momentum $\hat q^+= z_{\rm min}q^+$, with $z_{\rm min}\equiv {\rm min}(z,\, 1-z)$;
that is, $\hat Y=\ln(\hat q^+/q_0^+)$, with $q_0^+=Q_0^2/2P^-$ as before (since this is fully a proton scale).
The corresponding difference in the ``target'' rapidity is again obtained as $\hat \eta= \hat Y-\hat\rho$,
with $\hat\rho=\ln(\hat Q^2/Q_0^2)$ and $\hat Q^2=4/r^2$ the natural transverse resolution scale for the
dipole. Hence,
\beq
\hat \eta= \hat Y-\hat\rho= \ln\frac{\hat q^+}{q_0^+}-\ln\frac{\hat Q^2}{Q_0^2}=\ln\frac{z_{\rm min}q^+ P^-r^2}{2}
=\ln\frac{1}{x}+\ln\frac{z_{\rm min}Q^2r^2}{4}\,.
\eeq
We finally observe that \texttt{(i)} to the logarithmic accuracy of interest, we can approximate $z_{\rm min}
\simeq z(1-z)$, and \texttt{(ii)} the integral in  Eq.~\eqref{sigmalt} is controlled by values of $r$ such that
$4/r^2\sim z(1-z)Q^2$. Indeed,  in the perturbative regime where $T\propto r^2$, the integral over $r$ ``lives'' at
the largest possible values before it is eventually cut off by the exponential decrease of the
 modified Bessel functions (which exponentially vanish  when their arguments
become larger than one). This discussion implies $\hat\eta\simeq\eta$, as anticipated.

\bigskip
\bibliographystyle{utcaps}

\providecommand{\href}[2]{#2}\begingroup\raggedright\endgroup

\end{document}